# Topographic Modulation of Martian Near-Surface Winds: Insights from Perseverance Measurements and CFD Modeling in Jezero Crater

Yuhang Liu[1,2], Lei Zhang[1]*, Zhihao Shen[3], Peng Cao[4], Zhao Jiang[5], Jing Li[5], Jinhai Zhang[1]

[1]Key Laboratory of Deep Petroleum Intelligent Exploration and Development, Institute of Geology and Geophysics, Chinese Academy of Sciences, Beijing, China.

[2]College of Earth and Planetary Sciences, University of Chinese Academy of Sciences, Beijing, China

[3]Department of Civil and Environmental Engineering, The Hong Kong University of Science and Technology, Hong Kong SAR, China

[4]Faculty of Architecture, Civil and Transportation Engineering, Beijing University of Technology, Beijing, China

[5]Hexagon Manufacturing Intelligence (Qingdao) Co., Ltd., Qingdao, China

*Corresponding author: zhangl@mail.iggcas.ac.cn (L. Z.).

**Key Points:**

1. Microscale (9 km × 7km) wind fields in western Jezero Crater are characterized using Perseverance rover data.

2. Computational Fluid Dynamics (CFD) simulations demonstrate that local topography dominates the near-surface atmospheric flow patterns.

3. Crater-focused analyses reveal symmetric wind-direction deflection patterns between opposing inner walls and the crater floor.

## Abstract

Near-surface wind fields on Mars are profoundly modulated by complex topography, yet fine-scale wind field characteristics remain poorly resolved for key geomorphological units such as deltas, valleys, and impact craters, due to the spatial constraints of lander-based wind observations. To address this, we identified three dominant wind directions using measured near-surface wind data from the Perseverance rover at Jezero Crater and then integrated *in-situ* wind measurements with high-resolution numerical modeling. We established a high-resolution three-dimensional (3D) terrain model encompassing key local geomorphic units, including the delta, an impact crater, and nearby mesas, and performed Computational Fluid Dynamics (CFD) simulations under the above-mentioned three dominant wind directions. The results reveal a robust coupling mechanism between local topography and near-surface wind field structures. We demonstrate that wind speed is significantly enhanced over windward slopes but evidently attenuated within depressions and crater floors. Crucially, significant wind direction deflection angles were particularly evident in areas characterized by steeper slopes. For instance, wind flow exhibited a symmetrical deflection pattern along the opposing inner walls of the modeled impact crater, but stabilizing on the crater floor. Spatial comparisons indicate that wind deflection is most pronounced over steeper slopes, while sector-based distributions within the impact crater reveal a consistent symmetry between opposing wall and floor regions. These findings offer new and critical insights into the intimate connection between Martian surface aeolian erosion/deposition processes and local topographic evolution, which is vital for interpreting the sedimentary history of Jezero Crater.

**Plain Language Summary**

Using wind data observed by the Perseverance rover on Mars, we developed a 3D wind field model of western Jezero Crater. Our numerical simulations recreate daily wind patterns under different wind directions. We found that the local landscape, including the crater, delta, and mesas, strongly affects how the wind moves near the surface. Specifically, wind speeds were higher on windward slopes and lower in valleys and crater floors. Additionally, the wind direction tended to shift more sharply in steeper areas. This study provides valuable insights into how Mars' surface features influence wind patterns, which could help us better understand Martian climate and its long-term impact on the landscape.

# 1 Introduction

Aeolian processes are the dominant driver controlling near-surface deposition, erosion, and landscape evolution on Mars (Almeida et al., 2008; Bridges et al., 2017; Cutts & Smith, 1973; Greeley et al., 1992; Iversen & Greeley, 1985). These effects are visibly recorded across a wide range of landforms, from major dune fields and wind streaks to erosional depressions and depositional ridges (Lapôtre et al., 2016; Manent & El-Baz, 1986). However, these surficial features only capture the integrated, long-term effects of the aeolian regime (Greeley et al., 1992, 2001). The fundamental dynamic forcing mechanisms are the localized acceleration, deflection, and redistribution of boundary-layer airflow within complex terrain, which govern the formation and evolution of these landforms (Boazman et al., 2021; Chojnacki et al., 2019; Herkenhoff et al., 2023; Love et al., 2022). Despite their importance, systematic constraints on the spatial structure of this near-surface atmospheric flow remain insufficient, particularly in regions where steep topographic gradients coexist with localized sediment accumulations.

The western basin of Jezero Crater, characterized by the co-occurrence of deltaic deposits, dissected plateaus, secondary impact craters, and valleys, serves as a critical natural laboratory for understanding near-surface Martian aeolian dynamics (Holm-Alwmark et al., 2021; Mangold et al., 2020). While high-resolution orbital imagery reveals distinct patterns indicative of flow bifurcation and convergence in the resultant aeolian features (Day & Dorn, 2019; Wright et al., 2022), these interpretations are primarily derived from morphological evidence and currently lack rigorous validation from fine-scale dynamic modeling.

The Mars Environmental Dynamics Analyzer (MEDA) (Rodríguez-Manfredi et al., 2021) instrument aboard the Perseverance rover provides the highest temporal resolution *in-situ* observations of the Martian wind field to date, enabling the identification of the dominant wind patterns and diurnal variations within the Jezero landing site (Viúdez-Moreiras et al., 2022a, 2022b). Crucially, these measurements are inherently single-point-based, which limits their ability to capture topography-driven three–dimensional (3D) flow responses. Specifically, single-point measurements cannot directly reveal how winds are channeled, augmented, or recirculated across complex features such as the delta front escarpments, deep crater troughs, or within the interior of valleys. Consequently, to fully elucidate the directionality of aeolian erosion, sediment transport pathways, and the formation mechanisms of depositional centers, methods capable of resolving the near-surface airflow structure under realistic topographic forcing are critically required.

Martian microscale wind simulations are commonly conducted within a multi-scale nesting framework, in which global climate models (GCMs, 100–500 km resolution) (Rafkin et al., 2001) provide large-scale atmospheric forcing, mesoscale models provide kilometer-scale boundary conditions, and microscale solvers resolve the near-surface flow. This approach is indispensable in regions where *in-situ* meteorological observations are unavailable. However, when the horizontal grid spacing of a mesoscale model becomes comparable to the characteristic size of turbulent eddies in the Martian Planetary Boundary Layer (PBL)—typically several kilometers—the model enters the so-called 'terra incognita' or gray-zone regime (Newman et al., 2017; Senel et al., 2020; Shin & Hong, 2015). Within this transitional range, turbulence parameterizations originally designed for coarser resolutions may become inadequate and can lead to partial or double-representation of turbulent eddy effects. Such uncertainties in the mesoscale wind field directly propagate into the boundary conditions used for subsequent microscale simulations (Newman et al., 2017; Wyngaard, 2004).

In this paper, we directly use the observed MEDA wind speed and direction to define the inflow boundary conditions for the CFD simulations, thereby minimizing externally introduced uncertainties. In conjunction with high-resolution Digital Elevation Models (DEMs), we utilized the CFD software Cradle to construct 3D, microscale airflow simulations for the western Jezero Crater region to quantify the modulation of the near-surface wind field in Jezero by complex topography, and explore the dynamic linkages between the resulting wind field structure and aeolian erosion-deposition patterns.

## 2 Methods

### 2.1 Study area

Jezero Crater is located on the northwestern margin of the Isidis Basin on Mars (18.3°N~18.6°N, 77.2°E~77.5°E) with a diameter of ~45 km. This region hosts several geomorphically distinct units—including the prominent delta front escarpment, adjacent mesas, secondary impact craters, and local depressions (Figure 1a) —that exert strong topographic control on near-surface airflow. Orbiter and rover imagery (Fassett & Head, 2005; Goudge et al., 2015;

 indicate that the western inlet system delivered sediment into the crater, forming a well-preserved deltaic deposit and associated stratified sequences. Because aeolian erosion, sediment transport pathways, and depositional patterns are strongly modulated by local terrain geometry, resolving the fine-scale airflow over these features is critical for interpreting surface modification processes within Jezero Crater.

## 2.2 In-situ observation and representative wind scenario determination

Near-surface wind measurements were acquired by the MEDA instrument onboard the Perseverance rover at a sensor height of approximately 1.5 m above the surface. Wind data collected during the first 315 sols of the mission were analyzed, corresponding primarily to Martian northern hemisphere spring and summer, with solar longitude ($L_s$ 22°-153°). We analyzed the full wind speed and direction dataset to characterize the diurnal variability of the near-surface flow under relatively stable seasonal forcing (Figures 2a and 2b). The wind field observations indicate that the daytime airflows are predominantly ~~fom~~from southeast, while the wind direction shifts to the west at night, with generally weaker wind speeds. Furthermore, transitional east airflows are typically observed around sunset (Figures 2a). Wind rose diagrams constructed for the three seasonal sub-intervals within the analysis period (Figures 2c, 2d and 2e) show that the winds observed by the Perseverance rover exhibit only slight seasonal variation from spring to summer in the Northern Hemisphere, with no significant difference in dominant wind direction.

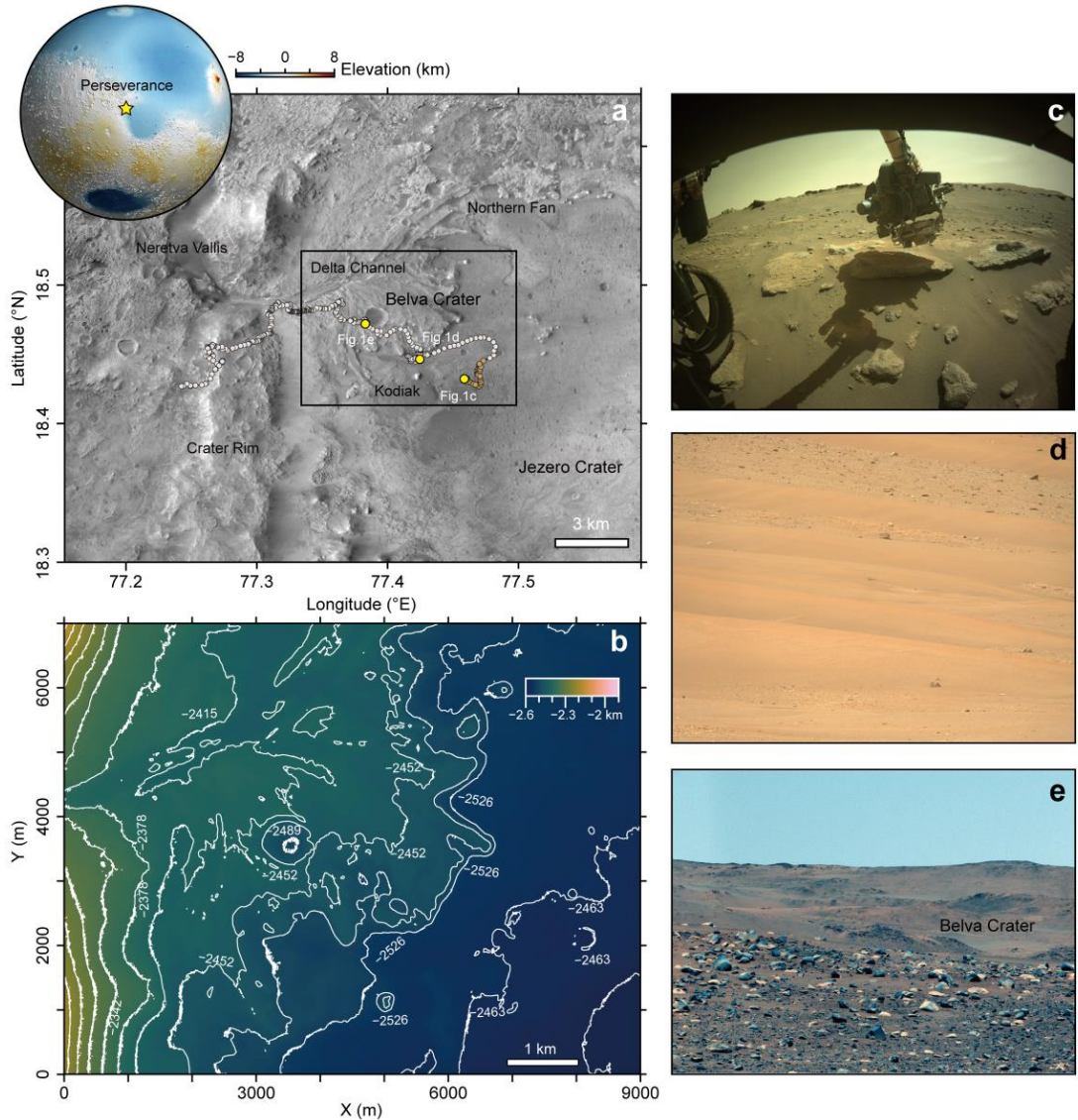

**Figure 1. Study area and in-situ observational images from the Perseverance rover in Jezero Crater. (a)** Location of Jezero Crater and the Perseverance rover landing region on Mars. Context Camera (CTX) image of the western Jezero Crater region, showing the delta deposits and surrounding terrain (image source: https://doi.org/10.5066/P9GV1ND3); white dots indicate the rover traverse, connected by gray lines. Yellow dots indicate the locations where the representative surface images in (c), (d), and (e) were acquired. The brown dots represent the position of the Perseverance rover during the first 315 Martian days before its wind sensor failed. **(b)** Elevation map of the study area. **(c)** Representative surface image acquired by the Hazcam at Sol 292. **(d–e)** Surface images acquired by the Mastcam-Z (Bell & Maki, 2021) at Sols 446 and 735, respectively, illustrating typical surface materials and terrain textures encountered along the traverse.

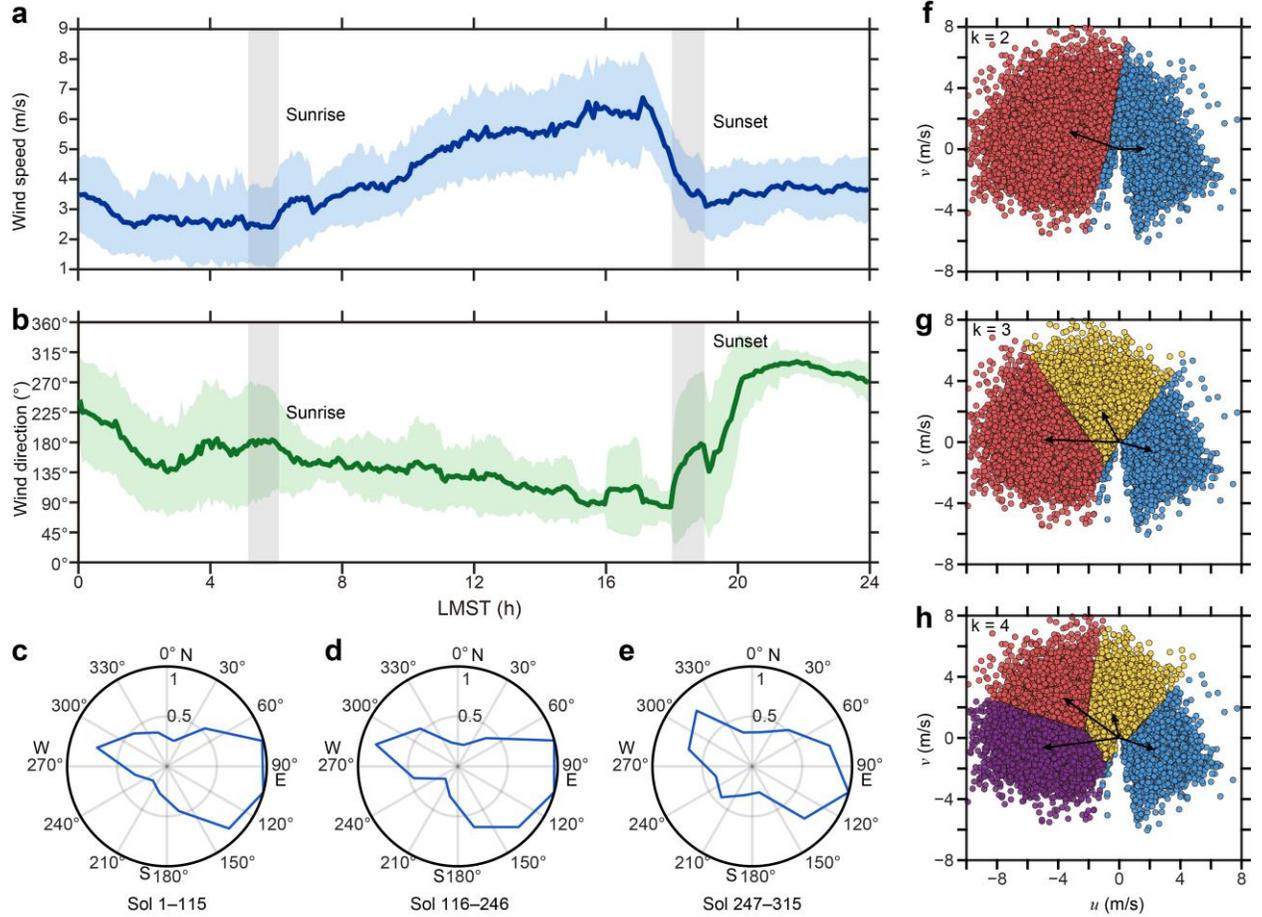

**Figure 2. Diurnal variability and dominant wind regimes at the Perseverance landing site.**
(**a–b**) Diurnal evolution of near-surface wind speed (top) and wind direction (bottom) measured by the MEDA instrument at the Perseverance landing region. Solid lines indicate mean values averaged over 10-min intervals, and shaded envelopes represent variability within the sampling window. Vertical dashed lines mark local sunrise and sunset. (**c–e**) Wind rose diagrams summarizing near-surface wind observations for three seasonal intervals: Sols 1–115, 116–246, and 247–315, respectively, illustrating the persistence of several preferred wind directions over seasonal timescales. (**f–h**) Results of $k$-means clustering applied to the 10-min averaged horizontal wind components ($u$, $v$), shown for $k$ = 2, 3, and 4. Colored points represent individual wind vectors, and black arrows indicate the mean wind direction of each cluster. The $k$ = 3 solution captures the dominant wind regimes used to define the representative inflow conditions for subsequent numerical simulations.

Representative wind regimes were identified using the $k$-means++ clustering algorithm (Ahmed et al., 2020; Bernardino et al., 2022; Yesilbudak, 2016). Tests with cluster numbers from $k$ = 2 to 8 showed monotonic trends in silhouette scores and sum of squared errors (SSE) values without a distinct optimum. Based on the dominant wind directions observed in the MEDA data and the corresponding wind-rose patterns (Figure 2), $k = 3$ was selected as the final clustering solution (Figure 2g). The three resulting clusters correspond to: Southeast wind (wind speed of 2.65 m/s with wind direction of 150.52°), east wind (wind speed of 5.25 m/s with wind direction of 91.91°), and west wind (wind speed of 2.62 m/s with wind direction of 285.25°). The MEDA measurements were interpreted as winds at a reference height ($Z_{ref}$)=1.5 m, forming the basis for constructing the vertical inflow wind profiles.

## 2.3 CFD Model and Governing Equations

Microscale wind simulations were conducted using the scSTREAM module of the Cradle CFD software, which solves the Reynolds-Averaged Navier–Stokes (RANS) equations with the Finite Volume Method. Pressure–velocity coupling is applied under an incompressible flow assumption, appropriate for the low Mach number conditions of the Martian near-surface boundary layer. Martian environmental parameters were specified as constant values representative of the MEDA observational season, including gravitational acceleration (3.7 m/s²), atmospheric density (0.016 kg/m³), and dynamic viscosity ($1.02 \times 10^{-5}$ Pa·s) (Munguira et al., 2023; Rodriguez-Manfredi et al., 2023). These values represent the ambient atmospheric state derived from *in-situ* measurements, where the direct mass-loading effect of suspended dust is considered negligible for the bulk flow properties under the simulated conditions. Additionally, the Coriolis force was neglected due to the microscale horizontal extent of the domain ($\sim 9\ km$); the calculated Rossby number ($Ro \approx 13.0$) indicates that inertial and topographic forcing dominate the flow dynamics over rotational effects at this scale (Holton & Hakim, 2013). Turbulence was modeled using the RNG $k$-$\varepsilon$ closure, which has demonstrated strong performance in separated and recirculating flows over complex terrain (Smith et al., 2017; Smyth & Hesp, 2019). The scSTREAM module of Cradle

CFD has been widely applied to microscale atmospheric flow simulations over complex terrain, including urban canopies, mountainous regions (Bhowmick et al., 2015; Gupta & Khare, 2021). Its finite-volume framework, together with robust turbulence closures, has demonstrated stable performance in resolving near-surface flow separation, recirculation, and topographically induced acceleration under low-Mach-number conditions. These characteristics make Cradle CFD well suited for investigating terrain-modulated wind fields at the spatial scales relevant to this study (Ouyang et al., 2024; Xuan et al., 2019).

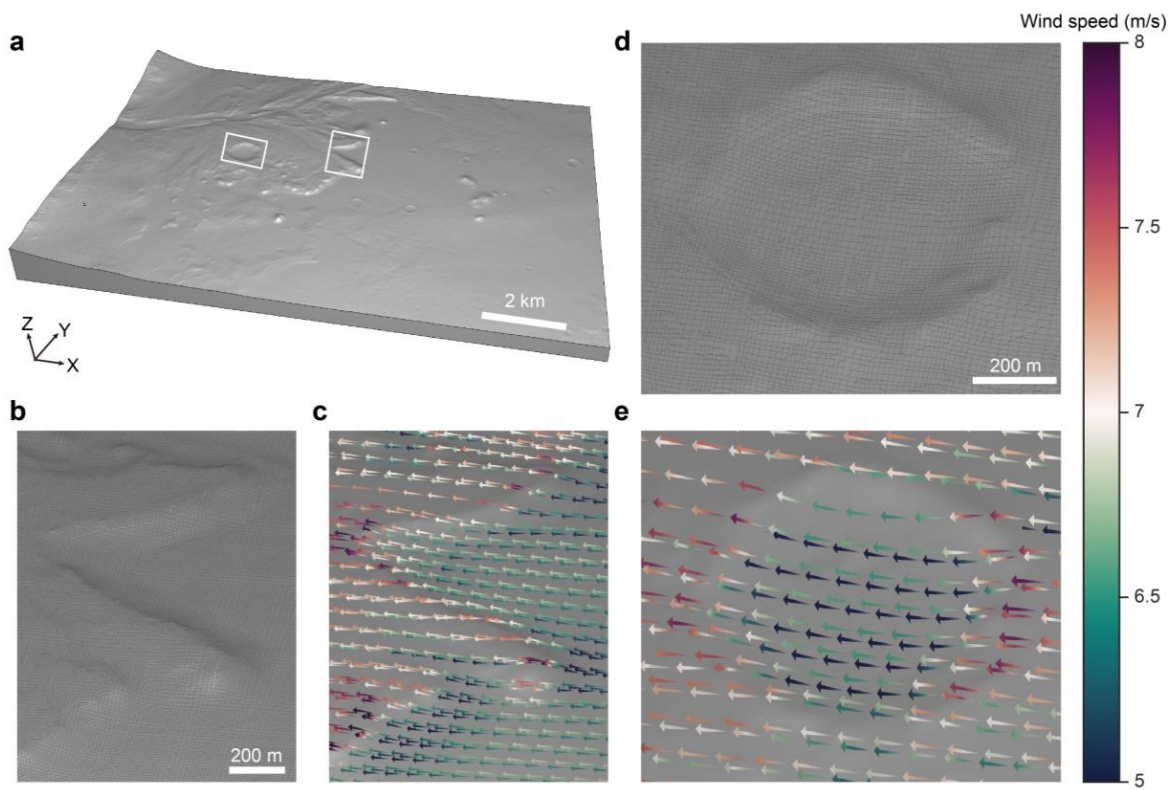

**Figure 3. Numerical model, mesh parameter setup and representative simulated streamlines.**
**(a)** Computational domain of CFD simulation. **(b)** Mesh construction of the delta deposition region **(c)** Streamlines distribution of the delta deposition region under east wind. **(d)** Mesh construction of the Delta Crater region. **(e)** Streamlines distribution of the Delta Crater region under east wind.

## 2.4 Computational domain and mesh

The topography was reconstructed using the Refubium dataset (Tao et al., 2023), originally from the CTX and HiRISE stereo imagery (0.5 meter spatial resolution). To optimize computational efficiency and ensure numerical stability without compromising essential geomorphic features, the high-resolution Digital Terrain Model (DTM) was systematically downsampled by a factor of 10, yielding a 5-meter spatial resolution for the mesh generation process. The computational domain spans 9.0 km × 7.0 km horizontally and extends 2.0 km vertically, placing the upper boundary at approximately five times the maximum local relief (~500 m) (Franke et al., 2011). The mesh consists of approximately 20 million cells, with a standard grid spacing of $15\ m \times 15\ m \times 5\ m$. In the near-surface region, the grid is refined to a minimal size of $2\ m \times 2\ m \times 2\ m$ to capture small-scale terrain-induced flow structures and satisfy boundary layer constraints. Further mesh coarsening was applied toward the domain boundaries and at higher altitudes to reduce computational cost while maintaining solution fidelity of interest.

## 2.5 Inflow construction

Because the scSTREAM module only permits specification of a power-law inflow profile (α), the vertical wind profiles could not be prescribed directly using a logarithmic formulation. To ensure consistency with the known aerodynamic roughness length at the Perseverance landing site ($z_0$=0.01 m) (Newman et al., 2021), we derived a power-law exponent α by fitting a neutral log-law wind profile over the 1–50 m height range, corresponding to the layer of primary interest in this study. The resulting exponent produces a power-law curve that accurately approximates the log-profile within the near-surface zone, while deviations at higher altitudes do not influence the shallow flow fields analyzed here. A limited sensitivity test (±0.2 variation in α ) showed that the spatial patterns of terrain-induced wind acceleration and direction deflection remained robust. We use three representative wind scenarios as boundary inputs (Figure 2g). Each representative wind scenario was simulated independently, with steady-state convergence achieved when normalized residuals decreased below $10^{-5}$ and monitoring points exhibited no further systematic drift.

## 3 Results

### 3.1 Wind speed and direction analyses

For all CFD simulation results, we display wind direction and speed of the 5 meters above the ground. Under the representative southeast inflow condition (wind speed of 2.65 m/s and wind direction of 150.52°), Figure 4 illustrates the near-surface wind field structure over the western depositional region of Jezero Crater and its relationship with local topography.

At the domain scale (Figure 4a), near-surface wind speeds are generally enhanced relative to the inflow speed, with most of the western depositional area exhibiting wind speeds between approximately 3 and 5m/s, corresponding to an aerodynamic enhancement of ~25%–80%. These enhanced wind speeds are primarily associated with areas of pronounced topographic relief, including the delta front, mesa margins, and impact crater rims. In contrast, the relatively flat eastern plain shows limited wind speed variability, with values remaining close to the background inflow speed, indicating a more uniform flow field in the absence of strong topographic perturbations. Notably, the simulation indicates that the rover was in a localized acceleration zone (Figure 4a). While this may result in higher absolute wind speeds in the model, using MEDA data is necessary to provide a physical anchor for the simulation. Since this research focuses on the spatial differentiation of the wind field, the absolute magnitude of the input does not change the spatial distribution patterns, which are primarily determined by the terrain geometry.

Over the alluvial fan and its upwind margins, wind speed distributions exhibit a clear topographic control. As the flow ascends the windward slope of the fan, wind speed increases from approximately 2–3 m/s in the lowlands to peak values of 4–5 m/s near the slope crest (Figures 4a and 4b). Immediately downstream of the crest, the flow undergoes rapid deceleration, forming localized low-speed zones (<1–2 m/s) within topographic depressions on the fan surface (Figure 4b). A similar wind speed evolution is observed across the Belva Crater. Under the inflow direction, airflow accelerates markedly along the windward crater rim, reaching localized maximum speeds of ~5 m/s. Upon entering the crater interior, wind speed rapidly decreases and stabilizes at approximately 2 m/s over the crater floor, forming a distinct low-speed region (Figure 4c). The flow subsequently re-accelerates after exiting the crater. These observations indicate that crater rims act as effective topographic obstacles, modulating near-surface wind speeds in a manner analogous to steep slopes. Within the channelized and fan-exit regions (Figure 4d), low wind speeds persist along topographic depressions, with values commonly remaining between 1.5 and 2.5 m/s over distances of up to ~4 km. In contrast, adjacent elevated terrain maintains relatively higher wind speeds.

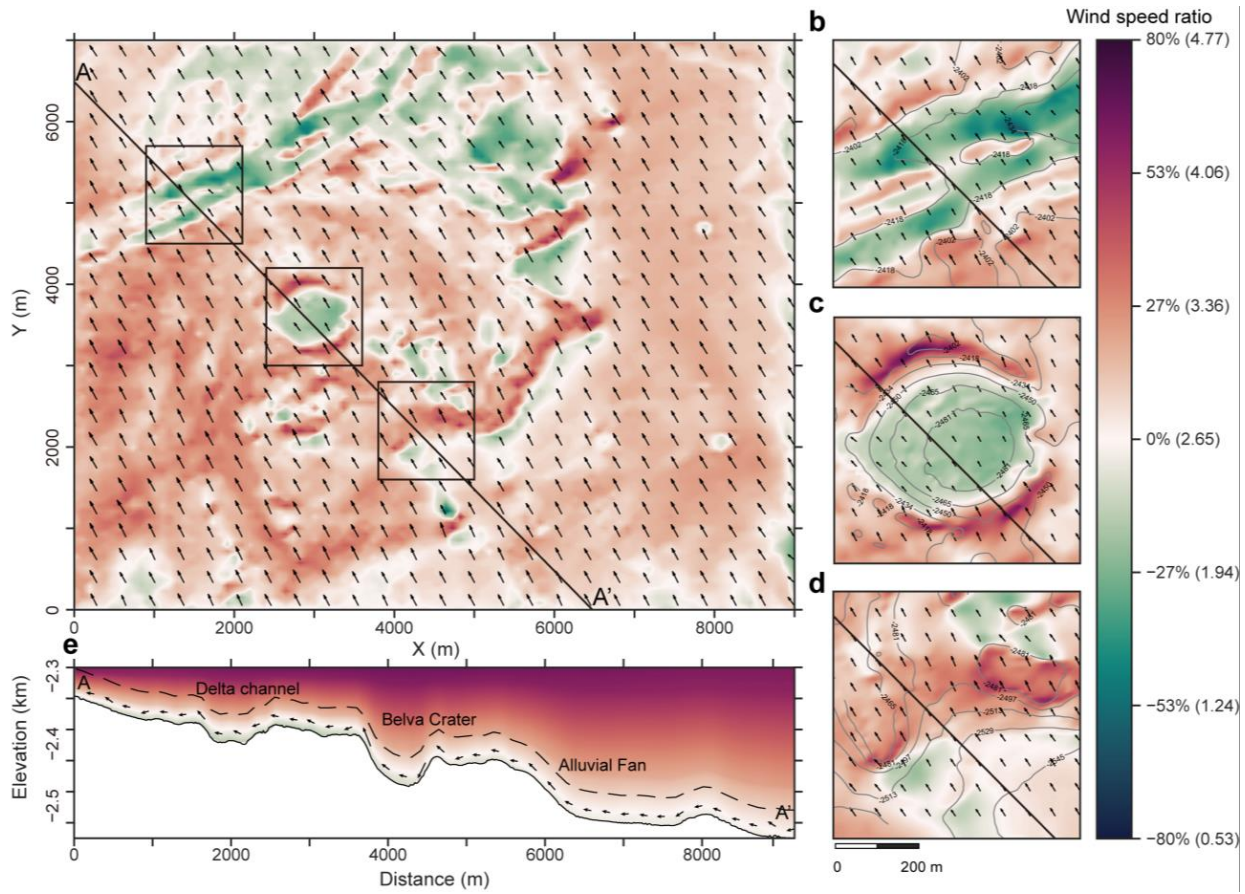

**Figure 4. CFD simulation results with southeast inflow: Near-surface wind speed ratio and flow direction. (a)** Domain-wide distribution; black boxes mark the regions enlarged in **b–d**. **(b–d)** Enlarged views of representative geomorphic units along transect A–A', illustrating wind field responses over the channel, impact crater, and alluvial fan region, respectively. **(e)** Vertical along-profile variation of wind speed ratio and terrain elevation extracted along transect A–A' (location shown in **a**). In this figure, the arrows indicate wind direction. Color shows the wind speed ratio (percentage) relative to the inflow speed for this case, with positive values indicating acceleration and negative values indicating deceleration. Grey contours represent elevation (m).

The vertical wind speed distribution extracted along the AA' transect (Figure 4e) demonstrates that the influence of topography is primarily confined to the shallow near-surface layer. Pronounced variations in wind speed associated with terrain occur within the low ~50 m above the surface, whereas at higher elevations the flow becomes progressively smoother and less sensitive to underlying topography. This indicates that the modulation of the wind field by alluvial fans, impact craters, and channels is largely restricted to the near-ground atmospheric layer.

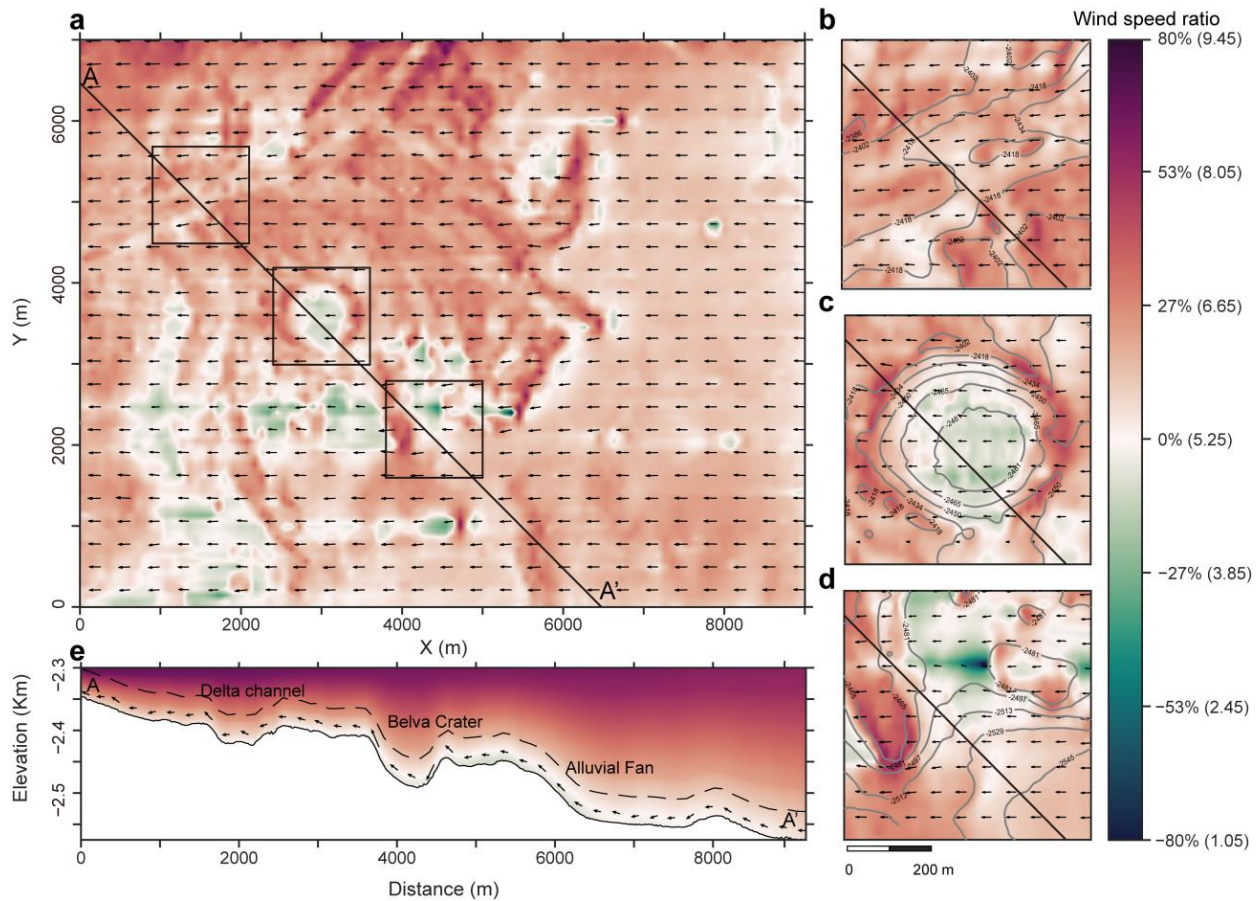

**Figure 5. CFD simulation results with east inflow: Near-surface wind speed ratio and flow direction. (a)** Domain-wide distribution; black boxes mark the regions enlarged in **b–d**. **(b–d)** Enlarged views of representative geomorphic units along transect A–A', illustrating wind field responses over the channel, impact crater, and alluvial fan region, respectively. **(e)** Vertical along-profile variation of wind speed ratio and terrain elevation extracted along transect A–A' (location shown in **a**). In this figure, the arrows indicate wind direction. Color shows the wind speed ratio (percentage) relative to the inflow speed for this case, with positive values indicating acceleration and negative values indicating deceleration. Grey contours represent elevation (m).

Figure 5 illustrates the simulated near-surface wind field under a higher-speed east inflow condition (wind speed of 5.25 m/s and wind direction of 91.91°). The resulting wind field shares several similarities with that obtained under the southeast inflow condition. At the domain scale (Figure 5a), wind speed enhancement remains widespread across the western depositional area of Jezero Crater. However, the spatial contrast between acceleration and deceleration zones is weaker than in the southeast wind scenario, and areas of enhanced wind speed are more uniformly distributed.

Within the paleo channel region (Figure 5b), low wind speed zones no longer persist continuously along linear topographic depressions. Instead, a more continuous acceleration pattern develops compared to the southeast simulation. This indicates that east winds tend to flow along shallow depressions, strengthening airflow within channels and producing wind speed enhancements exceeding 20%.

The deceleration patterns within the Belva crater are similar in both Figure 4c and Figure 5c, with low-speed zones mainly located on the crater floor. However, the acceleration areas along the outer rim show clear differences due to the different inflow directions. Under southeast inflow (Figure 4c), the wind speed increases primarily along the southeast diagonal of the rim. In contrast, under east inflow (Figure 5c), the acceleration pattern is symmetric along the east-west direction. Over the alluvial fan and adjacent mesas (Figure 5d), wind acceleration along windward slopes remains evident and is stronger than that observed under southeast inflow conditions. Downstream of the slope crests, deceleration zones are more scattered and spatially limited, which may be related to the alignment between wind direction and local slope orientation.

Although westerly winds originating from the Jezero Crater rim may be infrequent, we incorporated the wind data acquired by the Perseverance rover. Specifically, we focused on simulating the coherent westerly flow pattern observed by Perseverance using a western forcing condition (direction 285.2°, speed 2.62 m/s). As illustrated in Figure 6a, under this low input forcing, wind speeds throughout the study domain consistently ranged between 1 and 6 m/s. Localized speed maxima were distributed not only proximal to the cliff face but also across the northern sector of the sedimentary fan. Furthermore, significant high speed airflow was absent immediately along the downslope section of the cliff (Figure 6d); instead, a low wind speed channel persisted for approximately 1 km. Along Perseverance's early traverse route, localized acceleration was evident, enhancing the wind speed from approximately 2 m/s to 4 m/s. Importantly, this acceleration is not an isolated phenomenon; it is intrinsically linked to topographical obstacles downstream. Artuby Ridge, located near the Perseverance landing site, imposed a systematic modulation on the incident flow.

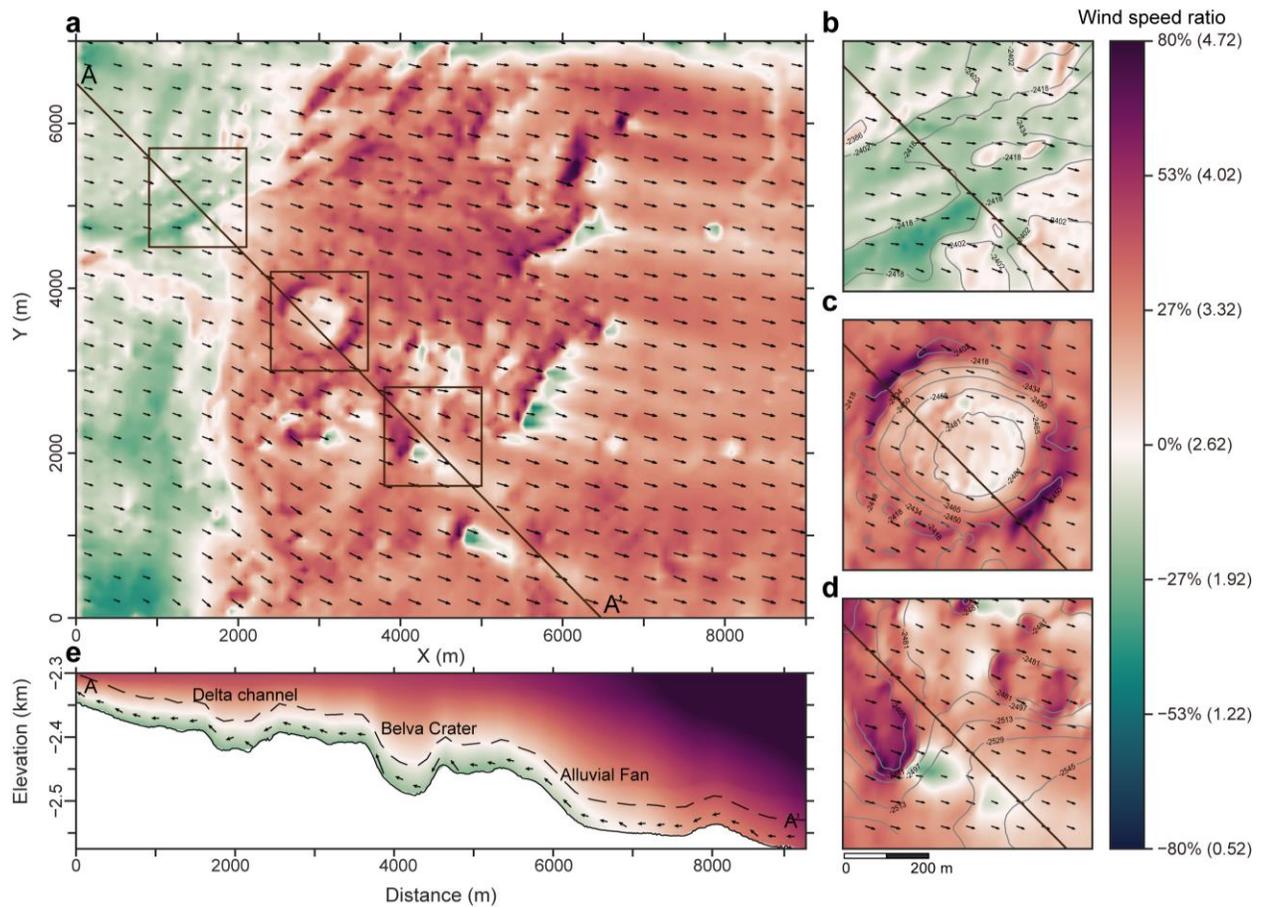

**Figure 6. CFD simulation results with west inflow: Near-surface wind speed ratio and flow direction. (a)** Domain-wide distribution; black boxes mark the regions enlarged in **b–d**. **(b–d)** Enlarged views of representative geomorphic units along transect A–A', illustrating wind field responses over the channel, impact crater, and alluvial fan region, respectively. **(e)** Vertical along-profile variation of wind speed ratio and terrain elevation extracted along transect A–A' (location shown in **a**). In this figure, the arrows indicate wind direction. Color shows the wind speed ratio (percentage) relative to the inflow speed for this case, with positive values indicating acceleration and negative values indicating deceleration. Grey contours represent elevation (m).

Under westerly inflow conditions (wind direction of approximately 285.25° and wind speed of 2.62 m/s), the simulated near-surface wind field shown in Figure 6 differs markedly from those obtained under southeast (Figure 4) and east (Figure 5) inflow conditions. At the regional scale (Figure 6a), the flow generally accelerates toward the eastern part of the domain, increasing from approximately 2 m/s over the western alluvial fan to 4–5 m/s near slope edges and mesa tops. At the same time, the areal extent of low-speed zones increases, mainly concentrated within the more rugged western portion of the alluvial fan. Additional small wind speed shadow zones also develop near the base of the delta deposits.

The wind field within the Belva Crater exhibits a complex response under westerly inflow conditions (Figure 6c). Wind speed is strongly enhanced along the windward crater wall, locally exceeding 4 m/s. In contrast to the other inflow cases, no clearly defined low-speed region develops over the crater floor, and the wind speed distribution is distinctly asymmetric. These results indicate that, under westerly inflow, the crater wind field is strongly controlled by the incoming wind direction.

The vertical wind speed distribution extracted along the AA′ transect (Figure 6e) shows that topographic modulation of wind speed is mainly confined to the lowest ~50 m above the surface, consistent with the other inflow conditions. However, compared to the southeast and east cases, the near-surface vertical wind speed gradient is more pronounced under west inflow, and the high-speed layer occupies a greater vertical extent on the eastern side of the transect. This suggests that terrain-induced vertical redistribution of momentum is stronger under westerly inflow conditions.

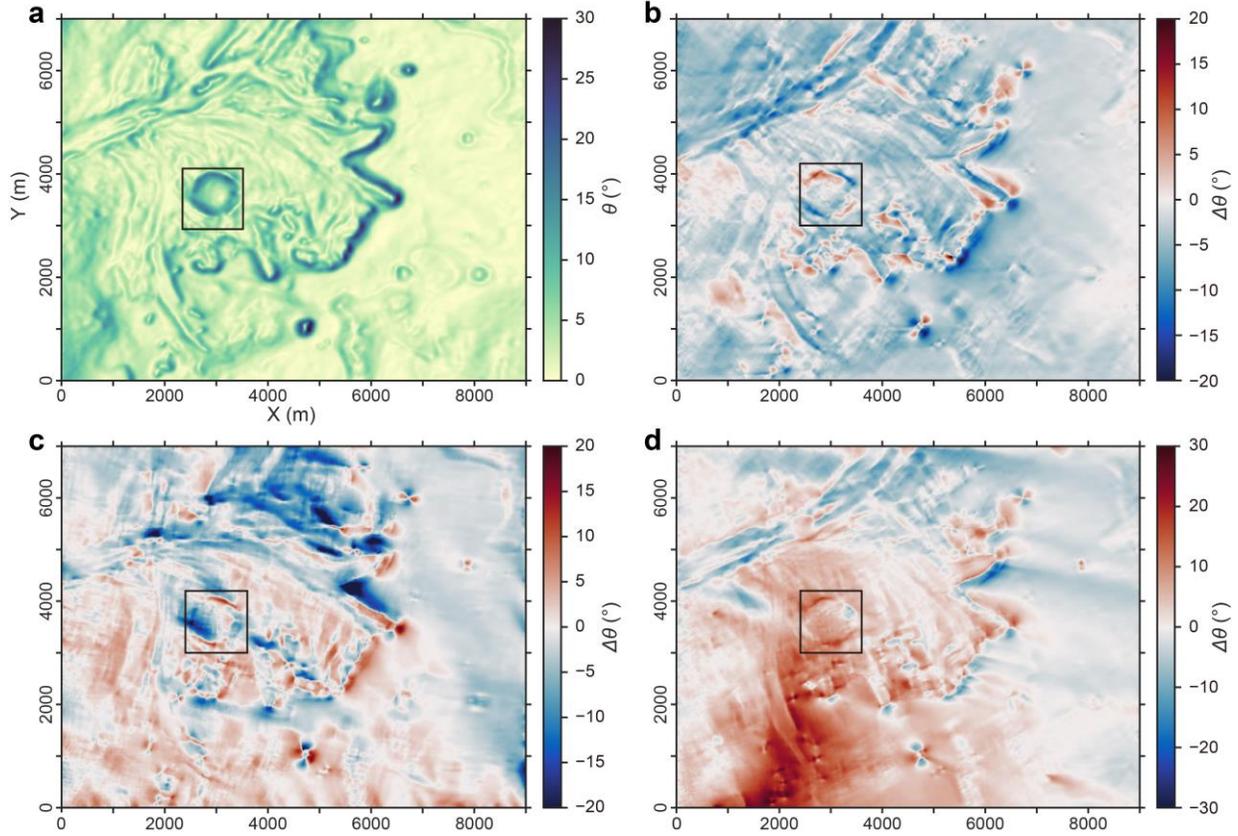

**Figure 7. Spatial relationship between surface slope and wind deflection under different inflow conditions. (a)** Spatial distribution of surface slope ($\theta$) in the study region, derived from the high-resolution digital terrain model. **(b–d)** show the corresponding near-surface wind deflection angle ($\Delta\theta$) under southeast, west, and east inflow conditions, respectively. The wind deflection angle $\Delta\theta$ is defined as the angular deviation of the simulated wind direction relative to the incident inflow direction, where positive values indicate clockwise deflection and negative values indicate counterclockwise deflection. Detailed distributions of $\Delta\theta$ in the Belva Crater (marked by black rectangles) are displayed in **Figure 8.**

### 3.2 Impact of Topography on wind direction deflection

By comparing the deflection angle between the incident and stabilized wind directions ($\Delta\theta$) with the regional slope field (Figure 7), the critical role of topographic gradient in controlling airflow redirection becomes evident. Under all three inflow conditions, wind direction changes across the Jezero delta consistently exhibit a tendency to adjust in accordance with local slope orientation, whereas pronounced wind deflection is primarily concentrated in regions characterized by strong topographic relief, such as impact craters and the alluvial fan.

Within the Belva Crater in particular, airflow adjacent to the crater walls is strongly constrained by lateral topographic slopes and is forced to undergo substantial directional turning. This turning effect is most pronounced in areas where the crater wall curvature increases and slopes become steeper. Along the inflow direction, wind deflection progressively intensifies along the upwind inner wall, reaches a maximum near the crater rim, and subsequently disperses and weakens on the leeward side. In contrast, the crater floor is characterized by relatively small slope and curvature, allowing the flow to traverse the region with minimal resistance and exhibiting only weak wind deflection close to background values. Spatially, two high-deflection bands are distributed concentrically along the crater wall, closely following the crater rim geometry, whereas wind deflection magnitudes ($|\Delta\theta|$) over the crater floor are markedly reduced and approach 0° (Figure 8).

This clear spatial contrast indicates the presence of two distinct flow states within the crater interior: a strongly deflected flow regime near the crater walls and a relatively stable wind field over the crater floor. In light of previous studies on flow over surface depressions, this wind field structure is highly consistent with the diffuser–confuser flow regime observed under low-speed conditions (Chen et al., 2023; Kovalenko et al., 2010; Tay et al., 2014). In this regime, the airflow remains fully attached while traversing the depression and does not develop large-scale flow separation or closed recirculation zones. The upstream half of the depression acts analogously to a diffuser, causing streamlines to converge toward the center, whereas the downstream half behaves as a confuser, forcing the streamlines to diverge outward. This characteristic "inward-then-outward" streamline curvature can be attributed to the reorientation of spanwise vortex lines induced by surface undulations. As the flow follows the concave surface, vortex lines originally aligned parallel to the surface are forced to bend downward and subsequently rise downstream, thereby inducing lateral velocity components near the surface. Without triggering flow separation, this process is sufficient to generate substantial wind direction deflection, with its intensity directly controlled by local slope and curvature.

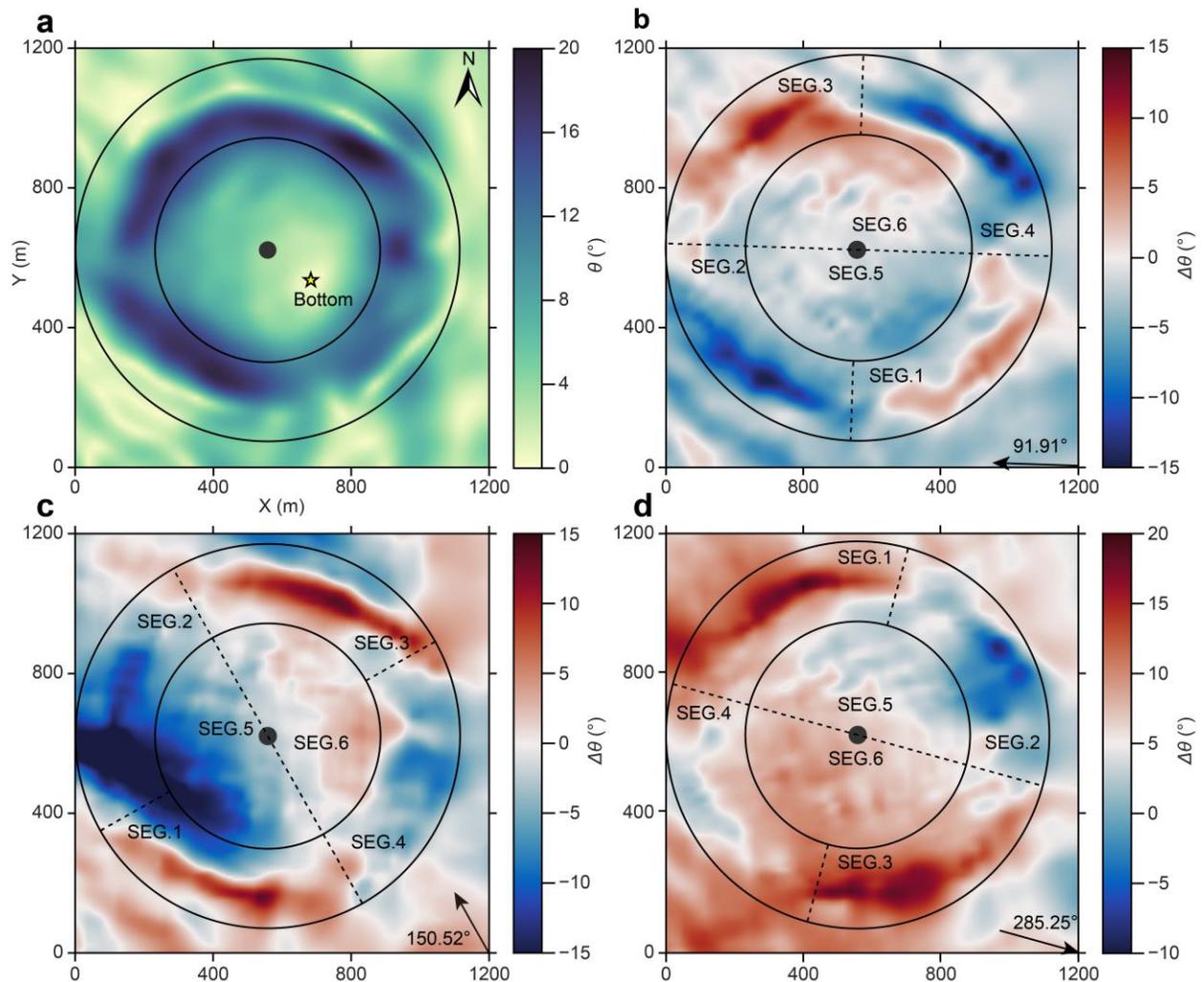

**Figure 8. Spatial relationship between surface slope and wind deflection within the Belva Crater under different inflow conditions. (a)** Surface slope distribution within the Belva Crater. **(b–d)** Spatial distribution of near-surface wind deflection angle $\Delta\theta$ within the crater under three different inflow conditions. The black arrows at the southeast corner of the subfigures indicate the wind inflow axis, and the two concentric solid lines represent the crater bottom ring and the crater

wall ring, respectively. Segments (SEG1–SEG6) indicate the azimuthal subdivisions used for localized comparison of wind deflection patterns. $\Delta\theta$ with positive values corresponding to clockwise deflection and negative values to counterclockwise deflection relative to the inflow direction.

The divergence and vorticity maps (Figure 9) provide a more detailed analysis of the flow field around Belva Crater. The divergence patterns show that the airflow undergoes a three-stage speed change: a slight acceleration as it approaches the outer rim, followed by a marked deceleration (inward) on the windward slope due to topographic obstruction. Upon exiting the crater, the flow then accelerates again (outward) on the leeward side. The vorticity distribution reflects the rotation of the airflow induced by the crater walls. In the southeast wind case (Figure 9e), the inner walls exhibit specific patterns of positive and negative vorticity, representing counterclockwise and clockwise rotations. Notably, the vorticity signs are inverted under the west wind case (Figure 9f). This opposite symmetry demonstrates that the local rotation and deflection of the wind are primarily controlled by the interaction between the inflow direction and the crater geometry. These observations confirm that the terrain systematically redirects the near-surface wind field without causing large-scale separation.

To further quantify the spatial variability of wind direction modulation within the crater, the Belva Crater interior was divided into six azimuthal segments (SEG.1–SEG.6) based on the inflow direction, and frequency distributions of near-surface wind deflection angles ($\Delta\theta$) were analyzed separately for the crater wall segments (SEG.1–SEG.4) and the crater floor segments (SEG.5–SEG.6) (Figure 10). Across all inflow conditions, the deflection angle distributions exhibit strong concentration, indicating that wind direction changes represent a stable response to geometric constraints rather than random fluctuations. Within the crater wall segments (SEG.1–SEG.4), deflection angle distributions are systematically shifted away from 0°, and their mean values vary coherently with inflow direction, reflecting persistent and directional turning of airflow along steep inner walls. In contrast, deflection angle distributions over the crater floor (SEG.5 and SEG.6) converge closely toward 0° and display significantly narrower spreads, indicating that airflow over low-slope, low-curvature surfaces remains largely aligned with the incident wind direction and experiences only weak background deflection.

Regarding spatial symmetry, azimuthally opposite sector pairs along the crater wall (SEG.1/SEG.3 and SEG.2/SEG.4) show similar deflection distributions under all three inflow scenarios, further supporting the interpretation that wind direction modulation is primarily governed by crater wall geometry rather than by localized flow instabilities. Nevertheless, differences in amplitude and spatial extent persist between clockwise and counterclockwise deflection bands, with the clockwise side relative to the inflow direction generally exhibiting slightly stronger and more extensive deflection.

This subtle asymmetry does not imply flow separation or vortex-dominated dynamics, but instead reflects an incomplete symmetry between inflow direction and crater geometry. Under fully attached flow conditions, the bending and reorientation of spanwise vortex lines over the concave surface induce lateral velocity components of slightly different magnitudes in different azimuthal sectors, leading to statistically asymmetric deflection distributions.

Overall, the statistical results presented in Figure 10 provide robust frequency-based support for the spatial patterns identified earlier: near-surface wind deflection within the Belva Crater primarily represents a stable, geometry-controlled response of attached flow, with both the magnitude and spatial distribution of deflection directly governed by variations in crater wall slope and curvature, rather than by flow separation or secondary vortical structures.

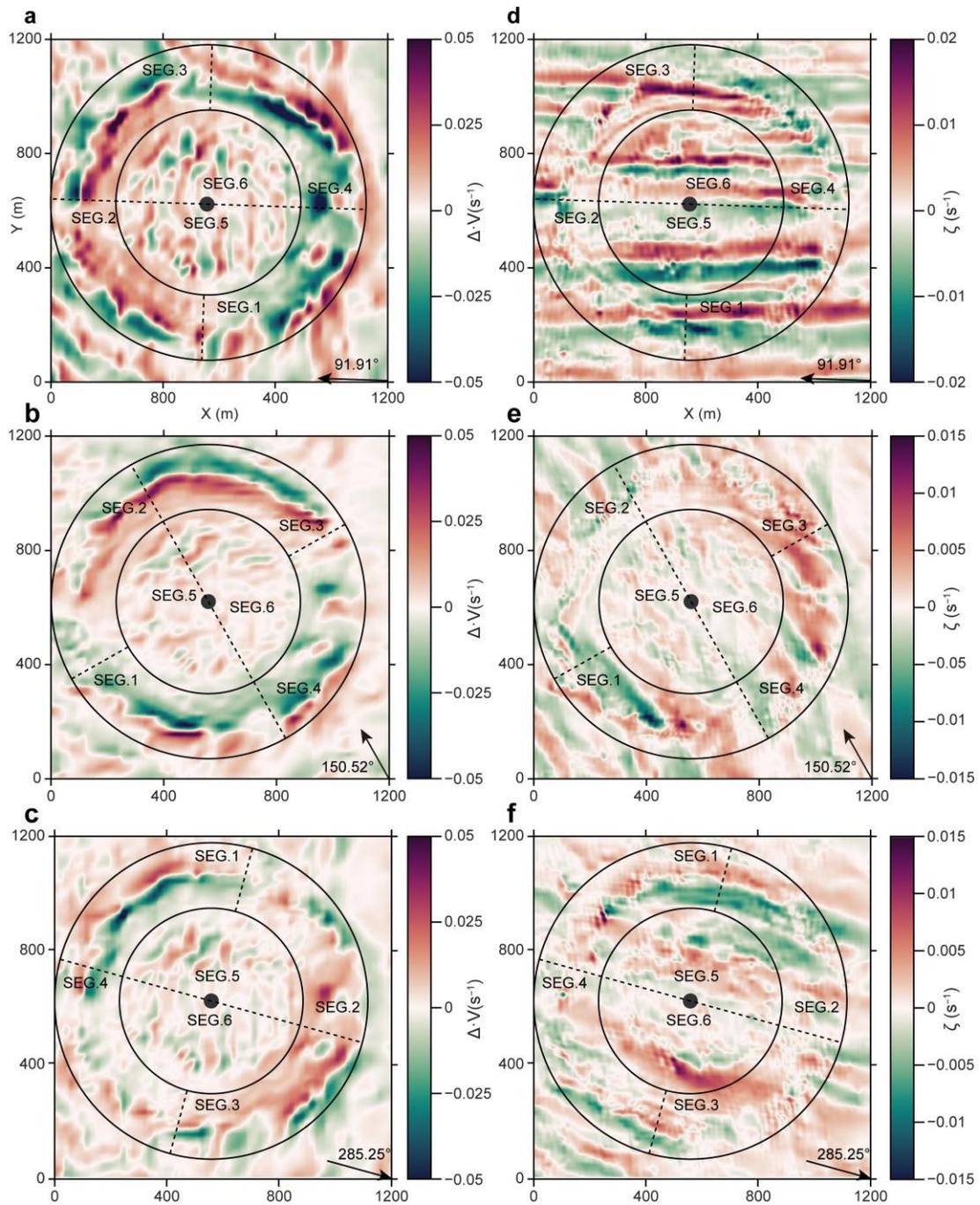

**Figure 9. Spatial distribution of divergence (left column) and vorticity (right column) for the near-surface wind field in Belva Crater. (a–c)** Divergence and convergence patterns: These panels illustrate the wind speed evolution as airflow traverses the crater. Green shading represents negative values (convergence/deceleration), and purple shading represents positive values (divergence/acceleration). **(d–f)** Vorticity distribution: These panels show the flow rotation

induced by the crater walls, where positive values indicate counter-clockwise rotation and negative values indicate clockwise rotation.

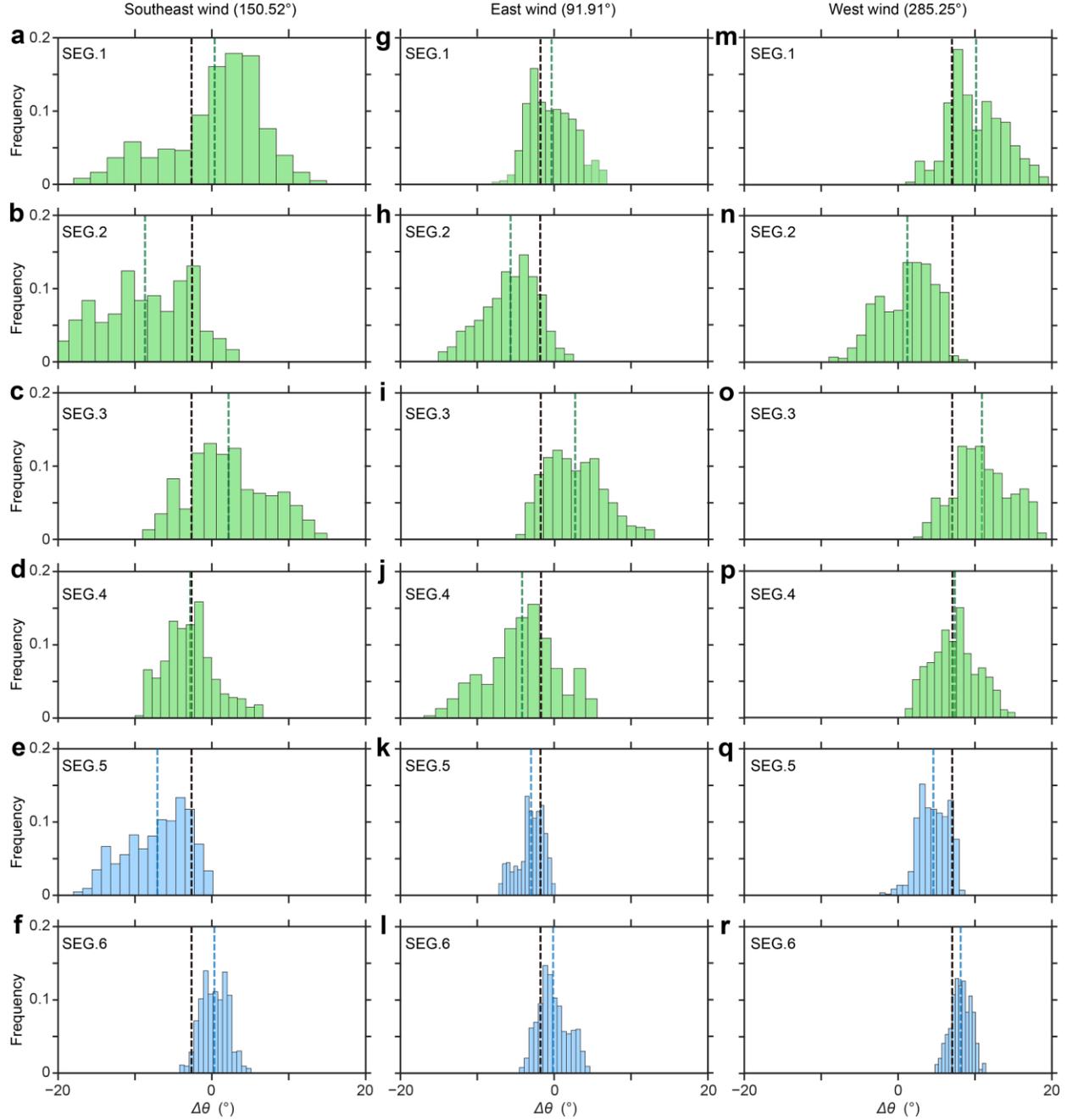

**Figure 10. Statistical frequency of wind deflection angles within different crater segments.**
**(a–d)** Frequency distributions of deflection angle ($\Delta\theta$) for the crater sidewall segments (SEG1–SEG4) with southeast inflow. Green dashed lines indicate the mean deflection angle for the corresponding sidewall segments, and the black dashed line denotes the crater-wide mean value.
**(e–f)** Frequency distributions of deflection angle ($\Delta\theta$) for the crater bottom segments (SEG5–SEG6) under southeasterly inflow. Blue dashed lines indicate the mean deflection angle for the crater bottom segments, and the black dashed line denotes the crater-wide mean value. **(g–l)** and **(m–r)**, same as **(a–f)**, but for east and west inflow conditions, respectively.

## 5 Discussions

This study used CFD simulations to characterize the spatial structure of the near-surface wind field at the meter resolution in the western region of Jezero Crater under the control of complex topography. The following content is worthy of discussion:

It is critical to emphasize that these model-elucidated characteristics reflect the steady-state, near-surface wind field structure under a given mean inflow condition, excluding high-frequency processes such as transient gusts or strong convective vortices. Although Martian aeolian erosion and dust events are often triggered by short-duration extreme wind speeds (Balme & Greeley, 2006; Hueso et al., 2023; Lemmon et al., 2022; Thomas & Gierasch, 1985), the topography-controlled mean flow structure likely determines the potential material transport pathways and spatial preference for deposition or erosion over longer timescales. For instance, persistent low-speed zones favor fine material deposition, while recurrent wind deflection and acceleration zones near slopes or crater walls may become preferential regions for sediment mobilization and transport during extreme events. Therefore, the spatial organization of the mean wind field constitutes the essential background framework for understanding the spatial distribution of extreme processes, rather than serving as the direct triggering mechanism.

This study employs measured MEDA wind data to directly constrain the inflow conditions of the microscale CFD model, a strategy that mitigates the uncertainties associated with downscaling from traditional mesoscale nesting approaches. This approach offers a distinct advantage in landing sites with high-quality in situ meteorological observations, ensuring that the simulations more directly reflect the true wind field response to local topography. The wind speeds recorded by MEDA are seasonally stable over the first 315 Sols, although they exhibit significant diurnal variability. This seasonal consistency is linked to the rover's traverse across the relatively flat units of the crater floor. In these level regions, the absence of major topographic obstacles results in a more uniform wind field, matching our simulation results for flat terrain. Since the rover had not yet entered rugged areas like the Belva crater rim or delta channels during this period, the current measurements do not reflect the large spatial variations in wind speed and direction identified in our model. Our simulations thus complement these localized observations by predicting the high spatial heterogeneity of wind regimes in complex terrains that the mission has yet to characterize *in-situ*. However, this method is inherently limited by not explicitly considering the temporal variability and multi-scale perturbation structure of the inflow conditions. Future work should prioritize assessing the impact of boundary condition uncertainty on microscale wind field structure, perhaps through integrating multiple wind scenarios or linking results with regional scale simulations.

Furthermore, by treating the surface as a rigid boundary, this modeling omits the feedback mechanisms of aeolian-surface coupling on surface roughness and flow structure. While this means the simulation results cannot directly predict erosion rates or deposition fluxes, they nevertheless provide robust dynamic constraints for identifying potential erosion and deposition hotspots. Subsequent research should incorporate particle tracking methods or simplified aeolian parameterization schemes to explore the inverse influence of particle transport on the near-surface flow structure.

Finally, while this study primarily characterizes the topographic modulation of the wind field, it does not explicitly account for the dynamic influences of thermal processes or atmospheric stability. On Mars, these thermal processes are inseparable from the terrain; the heating and cooling of slopes trigger significant slope winds (Montlaur et al., 2024), effectively acting as a thermal engine that drives near-surface circulation. Such synergy between topography and thermal forcing profoundly shapes the local microclimate. For instance, the persistent low-velocity zones within Belva Crater and other depressions identified in our models may facilitate cold-air pooling during Martian nights, while wind deflections along crater walls could affect moisture retention and diurnal temperature gradients. This reflects a dynamic coupling between surface energy balance and atmospheric motion: while diurnal surface heating and cooling cycles(Zhang et al., 2023; Zhang & Zhang, 2022) govern atmospheric stability and initiate slope winds, the resulting topographically deflected airflow modulates convective heat transfer and redistributes energy across the terrain, thereby reshaping local temperature gradients. While the present simulations focus on mechanically driven flow-topography interactions, future extensions that incorporate surface energy balance or thermal forcing, constrained by surface temperature or radiative observations, could further refine interpretations of aeolian processes and help explore their variability across diurnal and seasonal timescales in the Jezero region.

## 6 Conclusion

This study combines high–resolution digital terrain modeling with CFD simulations constrained by Perseverance's MEDA wind measurements to investigate the near–surface airflow organization in western Jezero Crater. Three representative wind regimes—approximately east, southeast, and west wind—were simulated to quantify the terrain–induced redistribution of wind speed and direction across the delta front, crater slopes, and valley networks. The simulation results show that Jezero's topographic relief produces robust patterns of flow acceleration over exposed slopes and crater rims, persistent low wind speed areas within depressions, and systematic wind–direction deflections along lateral valley walls. These structures, which arise independently of sediment feedbacks or thermal forcing, delineate potential pathways for preferential erosion and deposition under strong wind events. While the inflow conditions and physical processes considered here are simplified, the simulations provide essential baseline insights into how Jezero's geomorphology modulates winds relevant to dust lifting, sediment transport, and surface modification. Future work incorporating thermal stability, time–varying inflow, and coupled aeolian transport will further refine these interpretations and support the integration of dynamical modeling with observations of active surface changes.

**Conflict of Interest**

The authors declare no conflicts of interest relevant to this study.

**Acknowledgments**

This work is supported by the National Natural Science Foundation of China (Grants No. 42441810 and 42204178).

**Data Availability Statement**

Near-surface wind data used to define the inflow boundary conditions were obtained from the Mars Environmental Dynamics Analyzer (MEDA) instrument aboard the Perseverance rover as part of the Mars 2020 mission. MEDA data are archived at the Geosciences Node of NASA's Planetary Data System (PDS) and are available at https://pds-geosciences.wustl.edu/missions/mars2020/, with the MEDA bundle accessible via https://doi.org/10.17189/1522849. Mastcam-Z image products used for geomorphic context and surface characterization are also archived in the PDS and are available at https://doi.org/10.17189/1522843. The high-resolution DTM employed in this study was obtained from the Refubium repository (https://refubium.fuberlin.de/handle/fub188/38643), originally derived from HiRISE stereo imagery and reconstructed using deep-learning-based methods (Tao et al., 2023). CFD simulations were performed using the Cradle CFD (scSTREAM module; Hexagon/Software Cradle). The input files of a typical case for the CFD simulation are available online (https://github.com/zhanglei911/CFDJezero). The codes used in this study are available from the authors upon reasonable request.